\documentclass[12pt]{article}
\usepackage{graphicx}
\usepackage{amsmath}
\usepackage{amssymb}
\usepackage{caption2}
\begin{document}
\newcommand  {\ba} {\begin{eqnarray}}
\newcommand  {\be} {\begin{equation}}
\newcommand  {\ea} {\end{eqnarray}}
\newcommand  {\ee} {\end{equation}}
\renewcommand{\thefootnote}{\fnsymbol{footnote}}
\renewcommand{\figurename}{FIG.}
\renewcommand{\captionlabeldelim}{.~}

\vspace*{1cm}
\begin{center}
 {\Large\textbf{Fermion Masses and Flavor Mixing in A Supersymmetric $SO(10)$ Model}}

\vspace{1cm}
 \textbf{Wei-Min Yang\footnote{e-mail address: wmyang@ustc.edu.cn}$^{1}$ and Zhi-Gang Wang$^{2}$}

\vspace{0.3cm}
 \emph{CCAST(World Laboratory), P.O.Box 8730, Beijing 100080, P. R. China}\\
 $^1$\emph{Department of Modern Physics, University of Science and Technology of China, Hefei 230026, P. R. China}\\
 $^2$\emph{Department of Physics, North China Electric Power University, Baoding 071003, P. R. China}
\end{center}

\vspace{1cm}
 \noindent\textbf{Abstract}: we study fermion masses and flavor mixing in a supersymmetric
 $SO(10)$ model, where $\mathbf{10}$, $\mathbf{120}$ and $\mathbf{\overline{126}}$ Higgs
 multiplets have Yukawa couplings with matter multiplets and give masses to quarks and
 leptons through the breaking chain of a Pati-Salam group. This brings about that, at the
 GUT energy scale, the lepton mass matrices are related to the quark ones via several
 breaking parameters, and the small neutrino masses arise from a Type II see-saw mechanism.
 When evolving renormalization group equations for the fermion mass matrices from the GUT
 scale to the electroweak scale, in a specific parameter scenario, we show that the model
 can elegantly accommodate all observed values of masses and mixing for the quarks and
 leptons, especially, it's predictions for the bi-large mixing in the leptonic sector
 are very well in agreement with the current neutrino experimental data.

\vspace{1.0cm}

\vspace{0.3cm}
            \hspace*{2.35cm} physics

\newpage
 \noindent\textbf{I. Introduction}

\vspace{0.3cm}
 The flavor problem about fermion masses and mixing has been a open question in particle
 physics \cite{1}. In the Standard Model (SM), the fermion masses and mixing angles are
 completely arbitrary and neutrinos are massless, but recent experiments uniquely specify
 that neutrinos have small masses and bi-large leptonic mixing angles \cite{2}. The present
 global analysis give at $90\%$ C.L. \cite{3}
\begin{gather}
 \triangle m^{2}_{21}\approx7.1\times10^{-5}\;\mathrm{eV^{2}},\hspace{0.3cm}
 |\triangle m^{2}_{32}|\approx(1.3-3.0)\times10^{-3}\;\mathrm{eV^{2}}, \nonumber\\
 \tan^{2}\theta_{12}\approx0.4\,,\hspace{0.3cm} \sin^{2}2\theta_{23}>0.92\,,\hspace{0.3cm} \sin\theta_{13}<0.23\,,
\end{gather}
 which are distinctly different from large masses and small mixing angels in quark sector.
 Incorporating the diverse values of fermion masses and mixing into a theory with the small
 number of parameters is a large challenge for theoretical particle physicists. There have
 been many ideas proposed for the purpose \cite{4}, among others, supersymmetric (SUSY)
 models based on the grand unified theory (GUT) such as SUSY $SO(10)$ models are theoretically
 well-motivated extension of the SM and have received growing attention \cite{5}.

 SUSY $SO(10)$ models provide the most natural framework for generation of neutrino mass
 and realization of see-saw mechanism \cite{6}, in addition, the fermion mass hierarchies
 are also understood very well by appealing to additional family symmetries \cite{7a}.
 However, details of symmetry breaking chains and Higgs field structures plays a crucial
 role in specifying the model and determining the fermion masses and mixing angles \cite{5,7b}.
 One can generally obtain the correct quantum number for the SM particle content and
 implement the see-saw mechanism through the breaking chain of either the Pati-Salam group
 $SU(4)_{C}\otimes SU(2)_{L}\otimes SU(2)_{R}$ ($G_{422}$) or $SU(5)\otimes U(1)$ ($G_{51}$).
 The former breaks the left-right symmetry
 $SU(3)_{C}\otimes SU(2)_{L}\otimes SU(2)_{R}\otimes U(1)_{B-L}$ ($G_{3221}$) at the
 \emph{B}-\emph{L} energy scale and gives the right-handed neutrino masses by means of
 the renormalizable couplings of the fermion fields in $\mathbf{16}$ spinor representations
 with the Higgs multiplets in $\mathbf{\overline{126}}$ representation, in which the
 \emph{R}-parity is automatically conserved at all energy scales, whereas the latter makes
 use of the nonrenormalizable couplings of the matter spinor fields with the $\mathbf{16}$
 Higgs multiplets to generate neutrino masses. The different symmetry breaking chains give
 rise to different mass relations between quark sector and lepton sector. However, a
 successful GUT model is required to naturally account for the difference of masses and
 mixing between the quarks and the leptons. The minimal SUSY $SO(10)$ model \cite{8},
 where only one $\mathbf{10}$ and one $\mathbf{\overline{126}}$ Higgs multiplets have
 symmetric Yukawa couplings with matter multiplets, seems difficult to consistently
 incorporate the realistic neutrino oscillation parameters. It was recently realized
 that there are possibilities for reproducing observed fermion masses and mixing by
 considering new Higgs multiplets \cite{9}.

 In this works, we extend the minimal SUSY $SO(10)$ model by introducing antisymmetric
 couplings of one $\mathbf{120}$ Higgs multiplets with the matter fields. Supposing the
 $G_{422}$ breaking chain, the left-right symmetry $G_{3221}$ is broken by the $G_{422}$
 component $\mathbf{(10,1,3)}$ contained in the $\mathbf{\overline{126}}$ Higgs multiplets
 and large Majorana masses are given to the right-handed neutrinos. Below the \emph{B}-\emph{L}
 energy scale, this model is described as the minimal supersymmetric see-saw standard model
 with superfield couplings of the left-handed lepton doublets with the $SU(2)_{L}$ Higgs
 triplets, which are from the $(\mathbf{\overline{10},3,1})$ component of $\mathbf{\overline{126}}$
 and can give small Majorana masses to the left-handed neutrinos. The two Higgs doublets in
 the minimal supersymmetric standard model (MSSM) are linear combinations of the $SU(2)_{L}$
 doublet components from different $SO(10)$ representations of Higgses, which all contribute
 to electroweak symmetry breaking and give Dirac masses to all the fermions. The effective
 Majorana masses for the light left-handed neutrinos are generated through the Type II
 see-saw mechanism \cite{10}. Because of receiving contributions from different components
 of the $\mathbf{120}$ Higgs multiplets, the mass relation between the up-quark sector and
 the Dirac neutrino sector is different from the mass relation between the down-quark sector
 and the charged lepton sector. As we will see below, the characteristic structure of Higgs
 fields and symmetry breaking superpotential can lead to fit completely all the observed
 data of masses and mixing for the quarks and the leptons.

 The remainder of this paper is organized as follows. In Section II we outline the model
 and derive the GUT relation among the fermion mass matrices, and then the renormalization
 group equations (RGEs) of the mass matrices are introduced. In Sec. III, a detailed
 numerical analysis of the masses and mixing angles of the neutrinos is given in a
 specific parameter scenario satisfying the experimental constraints. Sec. IV is
 devoted to conclusions.

\vspace{1cm}
 \noindent\textbf{II. Model, Mass Matrices and Renormalization Group Evolution}

 \vspace{0.3cm}
 We consider the SUSY model based on the GUT gauge group $SO(10)$. The matter fields
 $\Psi^{16}$ in one $\mathbf{16}$ spinor representation of the $SO(10)$ group contain
 all the quarks and leptons as well as the right-handed neutrino of each generation.
 Higgs fields which can couple to spinor fermions at the renormalizable level include
 only the $H^{10}$, $H^{120}$, and $H^{\overline{126}}$ multiplets of the $\mathbf{10}$,
 $\mathbf{120}$, and $\mathbf{\overline{126}}$ representations under $SO(10)$, respectively.
 The gauge invariant Yukawa couplings superpotential are such as
 \ba W_{SO(10)}=y^{10}_{ij}\Psi^{16}_{i}H^{10}\Psi^{16}_{j}
                +y^{\overline{126}}_{ij}\Psi^{16}_{i}H^{\overline{126}}\Psi^{16}_{j}
                +y^{120}_{ij}\Psi^{16}_{i}H^{120}\Psi^{16}_{j}\,, \ea
 where $i,j$ are generation indices. By virtue of the gauge symmetry, the Yukawa couplings,
 $y^{10}_{ij}$ and $y^{\overline{126}}_{ij}$, are symmetric $3\times3$ matrices, while
 $y^{120}_{ij}$ is an antisymmetric one. In our model, all the Yukawa couplings are
 assumed to be real to keep the number of free parameters minimum.

 We suppose that the GUT gauge symmetry descends to
 $SU(3)_{C}\otimes SU(2)_{L}\otimes U(1)_{Y}$ ($G_{321}$) of the SM through two
 intermediate symmetries which are orderly the Pati-Salam subgroup $G_{422}$ and the
 left-right symmetry subgroup $G_{3221}$. At first, the $SO(10)$ symmetry is broken
 down to $G_{3221}$ by developing non-vanishing vacuum expectation values (VEVs)
 $\langle\mathbf{54}\rangle+\langle\mathbf{45}\rangle$ of two Higgs multiplets in Higgs
 superpotential. The subsequent breaking to $G_{321}$ is achieved by the $\mathbf{(10,1,3)}$
 component VEV of $H^{\overline{126}}$ in eq.(2), which is a singlet under $G_{321}$.
 In the framework of $G_{321}$, accordingly the Yukawa superpotential are rewritten as
\begin{alignat}{1}
 W_{G_{321}}=&\:\hat{Q}_{i}\left[y^{10}_{ij}H^{10}_{(1,2,2)_{u}}
             -\frac{1}{3}y^{\overline{126}}_{ij}H^{\overline{126}}_{(15,2,2)_{u}}
             +y^{120}_{ij}\left(H^{120}_{(1,2,2)_{u}}+\frac{1}{3}H^{120}_{(15,2,2)_{u}}\right)\right]\hat{u}^{c}_{j} \nonumber\\
            &+\hat{Q}_{i}\left[y^{10}_{ij}H^{10}_{(1,2,2)_{d}}
             -\frac{1}{3}y^{\overline{126}}_{ij}H^{\overline{126}}_{(15,2,2)_{d}}
             +y^{120}_{ij}\left(-H^{120}_{(1,2,2)_{d}}+\frac{1}{3}H^{120}_{(15,2,2)_{d}}\right)\right]\hat{d}^{c}_{j} \nonumber\\
            &+\hat{L}_{i}\left[y^{10}_{ij}H^{10}_{(1,2,2)_{d}}
             +y^{\overline{126}}_{ij}H^{\overline{126}}_{(15,2,2)_{d}}
             +y^{120}_{ij}\left(-H^{120}_{(1,2,2)_{d}}-H^{120}_{(15,2,2)_{d}}\right)\right]\hat{e}^{c}_{j} \nonumber\\
            &+\hat{L}_{i}\left[y^{10}_{ij}H^{10}_{(1,2,2)_{u}}
             +y^{\overline{126}}_{ij}H^{\overline{126}}_{(15,2,2)_{u}}
             +y^{120}_{ij}\left(H^{120}_{(1,2,2)_{u}}-H^{120}_{(15,2,2)_{u}}\right)\right]\hat{\nu}^{c}_{j} \nonumber\\
            &+\hat{L}_{i}\,y^{\overline{126}}_{ij}H^{\overline{126}}_{(\overline{10},3,1)}\hat{L}_{j}
             +\hat{\nu}^{c}_{i}\,y^{\overline{126}}_{ij}\left\langle H^{\overline{126}}_{(10,1,3)}\right\rangle\hat{\nu}^{c}_{j}\,,
\end{alignat}
 where as usually the notations $\hat{Q}, \hat{L}, \hat{u}, \hat{e}$, etc. denote the
 chiral superfields for the quarks and leptons, and the subscripts of various Higgs
 superfields show that they respectively originated from corresponding representations
 of $G_{422}$, the lowest indices $u,d$ stand for belonging to the up-type and down-type
 $SU(2)_{L}$ Higgs doublet of $G_{321}$, respectively. Note that a Clebsch-Gordon
 coefficient $(-3)$ is generated in the lepton sectors when the components
 $\mathbf{(15,2,2)}$ of $G_{422}$ are involved in the superpotential, which plays a
 crucial role in obtaining the Georgi-Jarlskog relations \cite{11}. The last second
 couplings will yield the left-handed neutrino Majorana masses after the neutral
 component of the $SU(2)_{L}$ Higgs triplet $H^{\overline{126}}_{(\overline{10},3,1)}$
 develops a VEV as small as the neutrino masses measured in experiments. The last term
 gives the right-handed neutrino Majorana masses, moreover, the VEV
 $\left\langle H^{\overline{126}}_{(10,1,3)}\right\rangle$
 should be close to the GUT energy scale so as to successfully implement the see-saw
 mechanism. At last, the electroweak symmetry breaking is accomplished by the electroweak
 scale VEVs of neutral components of these Higgs doublets in eq.(3), through which all
 the fermions acquire Dirac masses.

 From the point of view of phenomenology, omitting the heavy Higgs fields decoupling from
 the low energy (here we assume that a doublet-doublet Higgs mass splitting is realized),
 the two light Higgs doublets in the MSSM, whose VEVs are $v_{u},v_{d}$, are realistically
 identified with some appropriate linear combinations of the above up-type and down-type
 Higgs doublets, respectively. This corresponds that the up-type Higgs doublet VEVs are in
 proportion to $v_{u}$, while the down-type ones are relative to $v_{d}$. To sum up, the
 various Higgs field VEVs in eq.(3) are described by
\begin{gather}
 \left\langle H^{10}_{(1,2,2)_{u/d}}\right\rangle = c^{(1)}_{u/d}\,v_{u/d}\,,\hspace{0.3cm}
 \left\langle H^{\overline{126}}_{(15,2,2)_{u/d}}\right\rangle = c^{(2)}_{u/d}\,v_{u/d}\,,\nonumber\\
 \left\langle H^{120}_{(1,2,2)_{u/d}}\right\rangle = c^{(3)}_{u/d}\,v_{u/d}\,,\hspace{0.3cm}
 \left\langle H^{120}_{(15,2,2)_{u/d}}\right\rangle = c^{(4)}_{u/d}\,v_{u/d}\,,\nonumber\\
 \left\langle H^{\overline{126}}_{(\overline{10},3,1)}\right\rangle = c_{L}\frac{v^{2}}{M_{G}}\,,\hspace{0.3cm}
 \left\langle H^{\overline{126}}_{(10,1,3)}\right\rangle = c_{R}\,M_{G}\,,
\end{gather}
 where $v_{u}=\sin\beta\,v, v_{d}= \cos\beta\,v$, and $v=174$ GeV is the VEV of the
 Higgs field in the SM, $M_{G}$ is the GUT energy scale. The coefficients
 $c^{(k)}_{u/d}(k=1,\ldots,4)$ arising from the mixing of the Higgs doublets are in
 generally complex, through which $CP$-violating phases are introduced into the fermion
 mass matrices. The parameters $c_{L}, c_{R}$ are positive, by which the mass scales of
 the left-handed and right-handed neutrino are related to the electroweak and GUT energy
 scale $v,M_{G}$.

 After the above symmetry breaking, the lagrangian relevant to the fermion masses is now
 obtained as follows
\begin{alignat}{1}
 -\mathcal{L}_{\mathrm{mass}}=\:&\overline{u}_{Li}M^{u}_{ij}u_{Rj}+\overline{d}_{Li}M^{d}_{ij}d_{Rj}
                +\overline{e}_{Li}M^{e}_{ij}e_{Rj}+\overline{\nu}_{Li}M^{\nu}_{ij}\nu_{Rj} \nonumber\\
               &+\frac{1}{2}\overline{\nu}_{Li}M^{L}_{ij}\nu^{c}_{Lj}
                +\frac{1}{2}\overline{\nu}^{c}_{Ri}M^{R}_{ij}\nu_{Rj} + h.c. \,,
\end{alignat}
 where the Dirac mass matrices for the quarks and leptons, as well as the Majorana mass
 matrices for the left-handed and right-handed neutrinos are, respectively, given by
\begin{alignat}{2}
  M^{u}_{ij} &=v\sin\beta \left[c^{(1)}_{u}\,y^{10}_{ij}-\frac{1}{3}c^{(2)}_{u}\,y^{\overline{126}}_{ij}
              +\left(c^{(3)}_{u}+\frac{1}{3}c^{(4)}_{u}\right)y^{120}_{ij}\right], \nonumber\\
  M^{d}_{ij} &=v\cos\beta \left[c^{(1)}_{d}\,y^{10}_{ij}-\frac{1}{3}c^{(2)}_{d}\,y^{\overline{126}}_{ij}
              +\left(-c^{(3)}_{d}+\frac{1}{3}c^{(4)}_{d}\right)y^{120}_{ij}\right], \nonumber\\
  M^{e}_{ij} &=v\cos\beta \left[c^{(1)}_{d}\,y^{10}_{ij}+c^{(2)}_{d}\,y^{\overline{126}}_{ij}
              +\left(-c^{(3)}_{d}-c^{(4)}_{d}\right)y^{120}_{ij}\right], \nonumber\\
 M^{\nu}_{ij}&=v\sin\beta \left[c^{(1)}_{u}\,y^{10}_{ij}+c^{(2)}_{u}\,y^{\overline{126}}_{ij}
              +\left(c^{(3)}_{u}-c^{(4)}_{u}\right)y^{120}_{ij}\right], \nonumber\\
  M^{L}_{ij} &=\frac{2v^{2}}{M_{G}}\,c_{L}\,y^{\overline{126}}_{ij}\,, \nonumber\\
  M^{R}_{ij} &=2M_{G}\,c_{R}\,y^{\overline{126}}_{ij}\,.
\end{alignat}
 After integrating out the heavy right-handed neutrino fields decoupling from theory
 below the \emph{B}-\emph{L} breaking scale, the effective Majorana masses of the light
 left-handed neutrinos are realized by the type II see-saw mechanism as
 \ba M^{L}_{\mathrm{eff}}=M^{L}-M^{\nu}(M^{R})^{-1}(M^{\nu})^{T}. \ea
 All the fermion mass matrices in eq.(6) are characterized by two energy scales $v,M_{G}$,
 three positive breaking vacuum parameters $\tan\beta,c_{L},c_{R}$, three real Yukawa
 coupling matrices $y^{10}_{ij},y^{\overline{126}}_{ij},y^{120}_{ij}$, and eight complex
 mixing coefficients $c^{k}_{u},c^{k}_{d}$, but these parameters are not independent each
 other.

 Owing to only the three Yukawa coupling matrices, the fermion mass matrices in eq.(6)
 are, at the GUT scale, correlated each other. We define the following notations as
\begin{gather}
  M^{f}_{S,A}=\frac{M^{f}\pm(M^{f})^{T}}{2} \hspace{0.3cm} (f=u,d,e,\nu)\,, \nonumber\\
  r_{l}=\frac{c^{(l)}_{u}}{c^{(l)}_{d}}\tan\beta \hspace{0.3cm} (l=1,2,3)\,, \hspace{0.3cm}
  r_{4}=\frac{c^{(4)}_{u}}{c^{(3)}_{u}}\;, \hspace{0.3cm}
  r_{5}=\frac{c^{(4)}_{d}}{c^{(3)}_{d}}\;, \nonumber\\
  r_{L}=\frac{c_{L}}{c^{(2)}_{d}\cos\beta}\;, \hspace{0.3cm}
  r_{R}=\frac{c_{R}}{c^{(2)}_{d}\cos\beta}\;.
\end{gather}
 The matrices $M^{f}_{S,A}$ denotes the symmetric and antisymmetric parts of the fermion
 Dirac mass matrices, respectively. The dimensionless parameters $r_{l},r_{4},r_{5}$ are
 virtually some ratios of the Higgs doublet VEVs in eq.(4). By virtue of the special
 definitions, it is easy to derive the GUT relations among the quark and lepton mass
 matrices as follows
\begin{alignat}{2}
  M^{u}_{A}&=\frac{r_{3}(3+r_{4})}{-3+r_{5}}M^{d}_{A}\,, \nonumber\\
  M^{e}_{S}&=\frac{4}{r_{1}-r_{2}}M^{u}_{S}-\frac{3r_{1}+r_{2}}{r_{1}-r_{2}}M^{d}_{S}\,,\hspace{0.3cm}
               M^{e}_{A}=\frac{-3(1+r_{5})}{-3+r_{5}}M^{d}_{A}\,, \nonumber\\
  M^{\nu}_{S}&=\frac{r_{1}+3r_{2}}{r_{1}-r_{2}}M^{u}_{S}-\frac{4r_{1}r_{2}}{r_{1}-r_{2}}M^{d}_{S}\,,\hspace{0.3cm}
               M^{\nu}_{A}=\frac{3r_{3}(1-r_{4})}{-3+r_{5}}M^{d}_{A}\,, \nonumber\\
  M^{L} &=6r_{L}\left(\frac{v}{M_{G}}\right)\left[\frac{1}{r_{1}-r_{2}}M^{u}_{S}
         -\frac{r_{1}}{r_{1}-r_{2}}M^{d}_{S}\right], \nonumber\\
  M^{R} &=6r_{R}\left(\frac{M_{G}}{v}\right)\left[\frac{1}{r_{1}-r_{2}}M^{u}_{S}
         -\frac{r_{1}}{r_{1}-r_{2}}M^{d}_{S}\right].
\end{alignat}
 The symmetric parts of the fermion mass matrices are expressed as linear combinations of
 $M^{u}_{S}$ and $M^{d}_{S}$, while the antisymmetric parts of them are linearly dependent
 on only $M^{d}_{A}$. Various coefficients of the linear combinations are determined by the
 seven parameters $r_{1},\ldots,r_{5},r_{L},r_{R}$, through which the lepton mass matrices
 are predicted by the quark mass matrices.

 In the following discussion, we consider the model in a specific scenario. First,
 adopting a generation basis in which the up-type quark mass matrix $M^{u}$ is real
 and diagonal. Secondly, assuming that the down-type quark mass matrix $M^{d}$ is
 Hermitian. Namely, the quark mass matrices are described as
 \ba M^{u}=\mathrm{diag}\left(m^{0}_{u},m^{0}_{c},m^{0}_{t}\right), \hspace{0.3cm}
     M^{d}=U^{0}_{CKM}\:\mathrm{diag}\left(m^{0}_{d},m^{0}_{s},m^{0}_{b}\right)\left(U^{0}_{CKM}\right)^{\dagger}, \ea
 where $U^{0}_{CKM}$ is the CKM matrix with three mixing angles and one $CP$-phase in
 the quark sector \cite{12}, the superscript $0$ means that the quark masses and mixing
 are evaluated at the GUT scale (hereafter as such). As a result, it is very easy to know
 from eq.(6) that the relation $c^{(4)}_{u}=-3c^{(3)}_{u}$, and the coefficients
 $c^{(1)}_{u/d},c^{(2)}_{u/d}$ are real, while $c^{(3)}_{d}, c^{(4)}_{d}$ are pure
 imaginary. Lastly, making a supposition that the coefficients $c^{(3)}_{u},c^{(4)}_{u}$
 are also real. To collect these together, we can draw a conclusion that all the charged
 fermion mass matrices in eq.(6), $M^{u},M^{d},M^{e}$, are Hermitian, while all the neutral
 ones, $M^{\nu},M^{L},M^{R}$, are real. The $CP$-violating sources come from only the
 antisymmetric parts in the down-type quark and charged lepton sectors. Furthermore, It
 can be inferred by eq.(8) that $r_{4}=-3$, and $r_{1},r_{2},r_{5},r_{L},r_{R}$ are real
 (moreover, $r_{L}$ and $r_{R}$ have the same sign) but $r_{3}$ is a pure imaginary. In
 terms of eq.(9), now this six parameters in addition to the ten values of the quark
 masses and mixing in eq.(10) determine fully all the mass matrices in the lepton sector.

 On account of the Hermite matrix $M^{e}$ satisfying the constraint equations
\begin{gather}
  \mathrm{Tr}  M^{e}=m^{0}_{e}+m^{0}_{\mu}+m^{0}_{\tau}\,, \hspace{0.3cm}
  \mathrm{Det} M^{e}=m^{0}_{e}\cdot m^{0}_{\mu}\cdot m^{0}_{\tau}\,, \nonumber\\
  \mathrm{Tr} \left[(M^{e})^{2}\right]=(m^{0}_{e})^{2}+(m^{0}_{\mu})^{2}+(m^{0}_{\tau})^{2}\,,
\end{gather}
 the three parameters $r_{1},r_{2},r_{5}$ can actually be solved out via inputting the
 charged lepton mass eigenvalues. Although one can find two sets of real solutions, whose
 parts involved in $r_{1},r_{2}$ are the same except the parts $r_{5}$ are different, the
 two $M^{e}$ matrices determined respectively by them are only a transpose each other.
 Without loss of generality, we only consider the solution whose $r_{5}$ is relatively
 larger in the following analysis. Therefore, the three masses of the charged leptons fix
 completely the parameters $r_{1},r_{2},r_{5}$, sequentially determine the charged lepton
 mass matrix by eq.(9).

 Now the left parameters $r_{L},r_{R},r_{3}$ are restricted through fitting the masses and
 mixing in the neutrino sector. In terms of eq.(7), the effective left-handed neutrino
 Majorana mass matrix $M^{L}_{\mathrm{eff}}$ is a real symmetric, but only the two
 mass-squared differences are known from eq.(1), so we consider the following simultaneous
 equations
\begin{gather}
  \mathrm{Tr}  M^{L}_{\mathrm{eff}}=m^{0}_{1}+m^{0}_{2}+m^{0}_{3}\,, \hspace{0.3cm}
  \mathrm{Det} M^{L}_{\mathrm{eff}}=m^{0}_{1}\cdot m^{0}_{2}\cdot m^{0}_{3}\,, \nonumber\\
  \mathrm{Tr} \left[\left(M^{L}_{\mathrm{eff}}\right)^{2}\right]
              =(m^{0}_{1})^{2}+(m^{0}_{2})^{2}+(m^{0}_{3})^{2}\,, \nonumber\\
  \triangle^{0}m^{2}_{21}=(m^{0}_{2})^{2}-(m^{0}_{1})^{2}\,, \hspace{0.3cm}
  \triangle^{0}m^{2}_{32}=(m^{0}_{3})^{2}-(m^{0}_{2})^{2}\,.
\end{gather}
 Taking $\triangle^{0}m^{2}_{21},\triangle^{0}m^{2}_{32},r_{3}$ as input parameters,
 the values of $m^{0}_{1},m^{0}_{2},m^{0}_{3},r_{L},r_{R}$ are solved out. The normal
 (or inverted) hierarchy corresponds to $\triangle^{0}m^{2}_{32}>0$ (or $<0$). Note
 that four sets of real solutions can be found for eq.(12), but only two sets satisfy
 conditions that $r_{L},r_{R}$ have the same sign. In fact, the two sets of solutions
 are only contrary sign. Putting them into eq.(9) and using eq.(7), they actually give
 the same matrix $M^{L}_{\mathrm{eff}}$, so one can adopt either of them.

 To summarize the above scenario, there are in all sixteen independent input parameters
 evaluated at the GUT scale. They are six quark masses, three angles and one $CP$-phase in
 the quark CKM matrix, three charged lepton masses, two neutrino mass-squared differences
 and one free parameter $r_{3}$. Providing these parameter values, all the fermion mass
 matrices at the GUT scale are entirely determined. Furthermore, running the fermion mass
 matrices from the GUT scale to the electroweak scale according to the following introduced
 RGEs, we can compare our results with the experimental data of the fermion masses and mixing
 at the electroweak scale.

 Below the \emph{B}-\emph{L} scale, the right-handed neutrinos are decoupled by the see-saw
 mechanism. The model particle spectrum are identical to ones of the MSSM with the effective
 left-handed Majorana neutrinos. We introduce the Yukawa coupling squared matrices for the
 charged fermions as follows
\begin{gather}
  S_{u}=\frac{1+(\tan\beta)^{-2}}{v^{2}}(M^{u})^{\dagger}M^{u}, \hspace{0.3cm}
  S_{d}=\frac{1+(\tan\beta)^{2}}{v^{2}}(M^{d})^{\dagger}M^{d}, \nonumber\\
  S_{e}=\frac{1+(\tan\beta)^{2}}{v^{2}}(M^{e})^{\dagger}M^{e},
\end{gather}
 and then the one-loop close RGEs for these Yukawa coupling squared matrices, the effective
 neutrino mass matrix and the gauge coupling constants of $G_{321}$ are given by \cite{13}
\begin{alignat}{2}
 \frac{\mathrm{d}\alpha_{i}(\chi)}{\mathrm{d}\chi}&=\frac{b_{i}}{2\pi}\: \alpha^{2}_{i}\,,\hspace{0.3cm}(i=1,2,3)\\
 \frac{\mathrm{d}S_{f}(\chi)}{\mathrm{d}\chi}&=\frac{1}{16\pi^{2}}\:(S_{f}K_{f}+K_{f}S_{f})\,,\hspace{0.3cm}(f=u,d,e)\\
 \frac{\mathrm{d}M^{L}_{\mathrm{eff}}(\chi)}{\mathrm{d}\chi}&=\frac{1}{16\pi^{2}}
                 \left[M^{L}_{\mathrm{eff}}K_{\nu}+(K_{\nu})^{T}M^{L}_{\mathrm{eff}}\right],
\end{alignat}
 with
\begin{alignat}{2}
 K_{u}&=3S_{u}+S_{d}+\left[\mathrm{Tr}(3S_{u})-4\pi\left(\frac{13}{15}\alpha_{1}+3\alpha_{2}+\frac{16}{3}\alpha_{3}\right)\right]I,\nonumber\\
 K_{d}&=S_{u}+3S_{d}+\left[\mathrm{Tr}(3S_{d}+S_{e})-4\pi\left(\frac{7}{15}\alpha_{1}+3\alpha_{2}+\frac{16}{3}\alpha_{3}\right)\right]I,\nonumber\\
 K_{e}&=3S_{e}+\left[\mathrm{Tr}(3S_{d}+S_{e})-4\pi\left(\frac{9}{5}\alpha_{1}+3\alpha_{2}\right)\right]I,\nonumber\\
 K_{\nu}&=S_{e}+\left[\mathrm{Tr}(3S_{u})-4\pi\left(\frac{3}{5}\alpha_{1}+3\alpha_{2}\right)\right]I,
\end{alignat}
 where $\chi=\mathrm{ln}(Q/M_{G})$, $b_{i}=(33/5,1,-3)$, and $I$ is a $3\times3$ unit matrix.
 Note that the eq.(14) and eq.(15) close on themselves, in addition, we neglect the running
 effects for the right-handed Majorana neutrinos since their mass scale is close to the GUT
 scale. For given $\tan\beta$, the previous calculated mass matrices and one unified gauge
 coupling constant are taken as the input values at the GUT energy scale $Q_{GUT}=M_{G}$,
 consequently, we can solve the above RGEs numerically and achieve the values of
 $\alpha_{i}(\chi),S_{f}(\chi),M^{L}_{\mathrm{eff}}(\chi)$ at the electroweak energy scale
 $Q_{weak}=M_{Z}$. The electroweak scale fermion mass eigenvalues are subsequently obtained
 by diagonalizing the Yukawa coupling squared matrices and the effective neutrino mass matrix
 as follows
\begin{alignat}{2}
 U_{u}S_{u}(\chi_{w})U^{\dagger}_{u} &=\frac{1+(\tan\beta)^{-2}}{v^{2}}
  \mathrm{diag}\left(m^{2}_{u}(\chi_{w}),m^{2}_{c}(\chi_{w}),m^{2}_{t}(\chi_{w})\right),\nonumber\\
 U_{d}S_{d}(\chi_{w})U^{\dagger}_{d} &=\frac{1+(\tan\beta)^{2}}{v^{2}}
  \mathrm{diag}\left(m^{2}_{d}(\chi_{w}),m^{2}_{s}(\chi_{w}),m^{2}_{b}(\chi_{w})\right),\nonumber\\
 U_{e}S_{e}(\chi_{w})U^{\dagger}_{e} &=\frac{1+(\tan\beta)^{2}}{v^{2}}
  \mathrm{diag}\left(m^{2}_{e}(\chi_{w}),m^{2}_{\mu}(\chi_{w}),m^{2}_{\tau}(\chi_{w})\right),\nonumber\\
 U_{\nu}M^{L}_{\mathrm{eff}}(\chi_{w})U^{T}_{\nu} &=
  \mathrm{diag}\left(m_{1}(\chi_{w}),m_{2}(\chi_{w}),m_{3}(\chi_{w})\right),
\end{alignat}
 where $\chi_{w}=\mathrm{ln}(M_{Z}/M_{G})$ denotes that the fermion masses and mixing are
 evaluated at the electroweak scale. Accordingly, the quark and lepton mixing matrices are
 given by \cite{12,14}
 \ba U_{u}U^{\dagger}_{d}=U^{q}_{CKM}(\chi_{w})\,,\hspace{0.3cm}
     U_{e}U^{\dagger}_{\nu}=U^{l}_{CKM}(\chi_{w})\:\mathrm{diag}\left(e^{i\beta_{1}},e^{i\beta_{2}},0\right), \ea
 where $\beta_{1},\beta_{2}$ are two Majorana phases in the lepton mixing matrix.
 Finally, the mixing angles and $CP$-violating phases in the unitary matrices
 $U^{q,l}_{CKM}(\chi_{w})$ are worked out by the standard parameterization in ref. \cite{15}.

\vspace{1cm}
 \noindent\textbf{III. Numerical Results}

\vspace{0.3cm}
 In this section, we present numerical results of our model. First of all, we fix the
 electroweak scale, the GUT scale and the unified gauge coupling constants such as
 \ba M_{Z}=91.2\;\mathrm{GeV},\; M_{G}=3.1\times10^{16}\;\mathrm{GeV},\;
     \alpha_{1}(0)=\alpha_{2}(0)=\alpha_{3}(0)=0.0408\,. \ea
 According to eq.(14), three gauge coupling constants at the electroweak scale are found to be
 \ba \alpha_{1}(\chi_{w})\approx0.0168\,,\hspace{0.3cm} \alpha_{2}(\chi_{w})\approx0.0335\,,\hspace{0.3cm}
     \alpha_{3}(\chi_{w})\approx0.1172\,. \ea
 These are in accordance with the current experimental measures very well \cite{15}.

 Secondly, the Higgs VEV $v$ is also fixed to $v=174$ GeV, and the two representative values
 of $\tan\beta$ are taken as $\tan\beta=10$ and $\tan\beta=30$. In the case $\tan\beta=10(30)$
 (Note, hereafter as such, the values in the parenthesis are corresponding to the case
 $\tan\beta=30$), we input the quark and charged lepton masses as well as the quark mixing
 angles and $CP$-violating phase at the GUT scale as follows (in GeV)
\begin{gather}
 m^{0}_{u}=0.00106\,(0.00107),\hspace{0.3cm} m^{0}_{c}=0.307\,(0.312),\hspace{0.3cm}  m^{0}_{t}=136\,(143), \nonumber\\
 m^{0}_{d}=0.00125\,(0.00135),\hspace{0.3cm} m^{0}_{s}=0.0265\,(0.0286),\hspace{0.3cm} m^{0}_{b}=1.02\,(1.18), \nonumber\\
 m^{0}_{e}=0.000324\,(0.00035),\hspace{0.3cm} m^{0}_{\mu}=0.0685\,(0.074),\hspace{0.3cm} m^{0}_{\tau}=1.17\,(1.32), \nonumber\\
 s^{0}_{12}=0.2229\,(0.2229),\; s^{0}_{23}=0.0348\,(0.0339),\; s^{0}_{13}=0.003\,(0.003),\;
            \delta^{0}_{13}=59^{\circ}\,(59^{\circ})\,.
\end{gather}
 After using eq.(14) and eq.(15), the equivalent values of the above masses and mixing
 at the electroweak scale are solved out such that (in GeV)
\begin{gather}
 m_{u}(\chi_{w})\approx0.00233,\hspace{0.3cm} m_{c}(\chi_{w})\approx0.677,\hspace{0.3cm} m_{t}(\chi_{w})\approx181, \nonumber\\
 m_{d}(\chi_{w})\approx0.00439,\hspace{0.3cm} m_{s}(\chi_{w})\approx0.0929,\hspace{0.3cm} m_{b}(\chi_{w})\approx3.01, \nonumber\\
 m_{e}(\chi_{w})\approx0.000487,\hspace{0.3cm} m_{\mu}(\chi_{w})\approx0.103,\hspace{0.3cm} m_{\tau}(\chi_{w})\approx1.75, \nonumber\\
 s^{q}_{12}(\chi_{w})\approx0.2229,\; s^{q}_{23}(\chi_{w})\approx0.0412,\; s^{q}_{13}(\chi_{w})\approx0.0036,\;
   \delta^{q}_{13}(\chi_{w})\approx59^{\circ}.
\end{gather}
 They are completely consistent with the current status of the quark and charged lepton
 masses and the CKM matrix at the $M_{Z}$ scale \cite{15}.

 Lastly, we take the input values of the neutrino mass-squared differences at the GUT
 scale and the parameter $r_{3}$ as (remember that $r_{3}$ is a pure imaginary)
\begin{gather}
 \triangle^{0}m^{2}_{21}=1.39\,(1.50)\times10^{-4}\;\mathrm{eV^{2}},\;
 \triangle^{0}m^{2}_{32}=3.9\,(4.3)\times10^{-3}\;\mathrm{eV^{2}},\nonumber\\
  r_{3}=759i\,(1105i)\,.
\end{gather}
 Here, we only consider the case of the normal hierarchy, namely $\triangle^{0}m^{2}_{32}>0$.
 For the case of the inverted hierarchy, a similar analysis can also be performed. After the
 RGEs of eq.(14)--(16) running simultaneously, the output values of the masses and mixing
 for the effective left-handed neutrinos at the electroweak scale are found as follows (in eV)
\begin{gather}
 m_{1}\approx0.0084\,(0.0050),\; m_{2}\approx0.0119\,(0.0098),\; m_{3}\approx0.0463\,(0.0459),\nonumber\\
 \triangle m^{2}_{21}\approx7.1\,(7.1)\times10^{-5},\;
 \triangle m^{2}_{32}\approx2.0\,(2.0)\times10^{-3},\nonumber\\
 \sin^{2}2\theta_{12}\approx0.816\,(0.816),\;
 \sin^{2}2\theta_{23}\approx0.960\,(0.948),\;
 \sin^{2}2\theta_{13}\approx0.154\,(0.082),\nonumber\\
 \tan^{2}\theta_{12}\approx0.40\,(0.40),\; \sin\theta_{13}\approx0.20\,(0.14),\nonumber\\
 \delta^{l}_{13}\approx -0.041\pi\,(-0.039\pi),\;
 \beta_{1}\approx 1.045\pi\,(0.045\pi),\; \beta_{2}\approx 0.495\pi\,(1.495\pi)\,.
\end{gather}
 The above results are excellently in agreement with the recent neutrino oscillation
 experimental data in eq.(1). The three neutrino masses are all within experimental limits.
 The values of $m_{2}$ and $m_{3}$ are close to $0.01$ eV and $0.046$ eV, respectively,
 but $m_{1}$ varies with $\tan\beta$ about range $(0.005-0.008)$ eV. The two mass-squared
 differences and the three mixing angles are all at the center values of the allowed
 regions. The leptonic $CP$-violating phase $\delta^{l}_{13}$ is approximate to $-0.04\pi$
 ($\approx -7.2^{\circ}$), which is small very much in comparison with the quark one. One of
 the two Majorana phases is about $\pi$ (or $0$), while the other is $\pi/2$ (or $3\pi/2$)
 or so. In addition, the right-handed neutrino masses are obtained straightforward by
 diagonalizing the mass matrix $M^{R}$ in eq.(9) such that (in GeV)
\begin{gather}
 M_{1}\approx9.1\times10^{12}\,(1.1\times10^{14}),\;
 M_{2}\approx1.7\times10^{14}\,(3.4\times10^{14}),\nonumber\\
 M_{3}\approx1.7\times10^{15}\,(4.8\times10^{15})\,.
\end{gather}
 They are close to the GUT scale except that the $M_{1}$ value is slightly low for the
 case $\tan\beta=10$, which are just expected in the previous sections.

 Now we illustrate the numerical analysis in detail. In Figs.~1 and 2 the three leptonic
 mixing angles $\sin^{2}2\theta_{ij}$ ($ij=12,23,13$) are shown as functions of the model
 parameter $r_{3}/i$ for the case $\tan\beta=10$ and $\tan\beta=30$, respectively. In the
 case $\tan\beta=10\,(30)$, we fix the input
 $\triangle^{0}m^{2}_{21}=1.39\,(1.50)\times10^{-4}\;\mathrm{eV^{2}}$ at the GUT scale so
 that $\triangle m^{2}_{21}\approx7.1\times10^{-5}\;\mathrm{eV^{2}}$ is always held at the
 electroweak scale. If the value of $\triangle^{0}m^{2}_{32}$ at the GUT scale is taken as
 $\triangle^{0}m^{2}_{32}=3.9\,(4.3)\times10^{-3}\;\mathrm{eV^{2}}$, which leads to
 $\triangle m^{2}_{32}\approx2.0\times10^{-3}\;\mathrm{eV^{2}}$ at the electroweak scale,
 this case is depicted by the red curves in each figure. If
 $\triangle^{0}m^{2}_{32}=5.9\,(6.4)\times10^{-3}\;\mathrm{eV^{2}}$, the value of
 $\triangle m^{2}_{32}$ is accordingly increased to
 $\triangle m^{2}_{32}\approx3.0\times10^{-3}\;\mathrm{eV^{2}}$, this case corresponds to
 the blue curves in each figure. The other input parameters are given by eq.(20) and (22).
\begin{figure}
 \centering
 \includegraphics[totalheight=7cm]{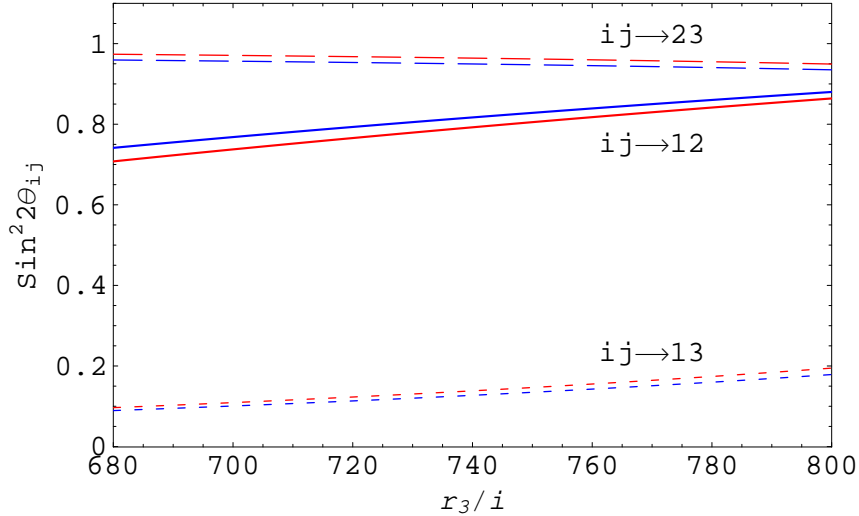}
 \caption{For $\tan\beta=10$, the three leptonic mixing angles as functions
  of the parameter $r_{3}/i$\,. The red and blue curves correspond to
  $\triangle m^{2}_{32}\approx 2.0\times 10^{-3}\,\mathrm{eV^{2}}$ and
  $3.0\times 10^{-3}\,\mathrm{eV^{2}}$, respectively, and they all hold
  $\triangle m^{2}_{12}\approx 7.1\times 10^{-5}\,\mathrm{eV^{2}}$.}
\end{figure}
\begin{figure}
 \centering
 \includegraphics[totalheight=7cm]{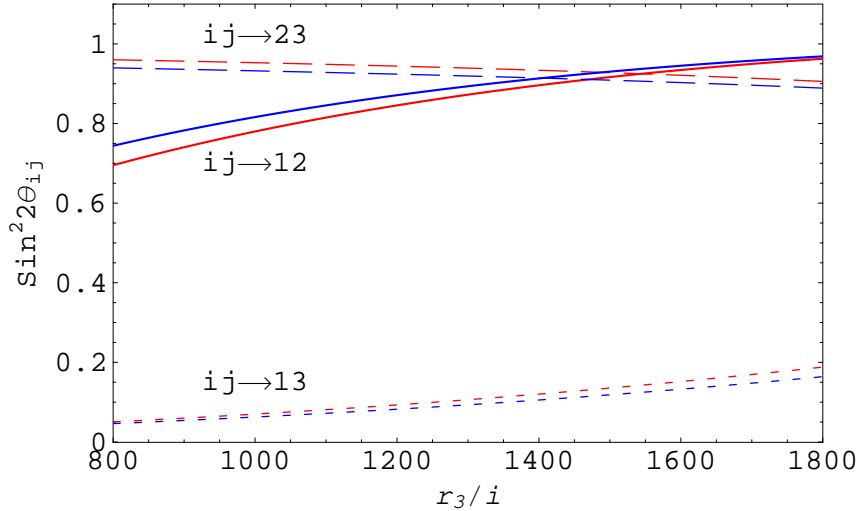}
 \caption{For $\tan\beta=30$, the three leptonic mixing angles as functions
  of the parameter $r_{3}/i$\,. The red and blue curves correspond to
  $\triangle m^{2}_{32}\approx 2.0\times 10^{-3}\,\mathrm{eV^{2}}$ and
  $3.0\times 10^{-3}\,\mathrm{eV^{2}}$, respectively, and they all hold
  $\triangle m^{2}_{12}\approx 7.1\times 10^{-5}\,\mathrm{eV^{2}}$.}
\end{figure}

 The graphs show explicitly the bi-large mixing characteristic of the lepton flavor mixing,
 namely, a very large solar angle $\sin^{2}2\theta_{12}\sim(0.7-0.9)$, a nearly maximal
 atmospheric angle $\sin^{2}2\theta_{23}\sim(0.9-1)$, and a very small CHOOZ angle
 $\sin^{2}2\theta_{13}\sim(0.05-0.2)$. As the value of $r_{3}/i$ is increased in the
 displayed range, the atmospheric angle $\sin^{2}2\theta_{23}$ is almost horizontal and
 always hold the near maximum. The solar and CHOOZ angles go up slowly, but they still
 lie in the experimental bounds. The best fit for the experimental data is at
 $r_{3}/i\sim760\,(1100)$ for the case $\tan\beta=10\,(30)$. In addition, when the
 electroweak scale $\triangle m^{2}_{23}$ varies from $2.0\times10^{-3}\,\mathrm{eV^{2}}$
 to $3.0\times10^{-3}\,\mathrm{eV^{2}}$, the plots only shift finely without changing the
 overall trends. Finally, it is worth to point out that $r_{3}/i$ varying has almost no
 effect on the values of $\triangle m^{2}_{12}$ and $\triangle m^{2}_{32}$ at the
 electroweak scale, which are principally determined by the GUT scale
 $\triangle^{0}m^{2}_{12}$ and $\triangle^{0}m^{2}_{32}$, respectively. In short, our
 model predict naturally the bi-large mixing structure, moreover, the three mixing angles
 and two mass-squared differences of the neutrino oscillations are realized simultaneously
 for certain parameter regions.

 \vspace{1cm}
 \noindent\textbf{IV. Conclusions}

\vspace{0.3cm}
 In summary, we have discussed the fermion masses and flavor mixing within the framework
 of the SUSY $SO(10)$ model. In the GUT model, we introduce the antisymmetric couplings
 of the $\mathbf{120}$ Higgs with the matter fields besides the symmetric couplings of
 the $\mathbf{10}$ and $\mathbf{\overline{126}}$ Higgs in the minimal SUSY $SO(10)$ model.
 The new Yukawa couplings and the left-right symmetry breaking chain not only bring about
 that the quark and lepton mass matrices are closely related to each other, but also lead
 to elegant explanation for the difference between the large flavor mixing for the leptons
 and the small flavor mixing for the quarks. After the renormalization group evolution
 from the GUT scale to the electroweak scale, all the current experimental data for the
 fermion masses and mixing as well as the gauge couplings are reproduced correctly in our
 model parameter scenario. In particular, the bi-large mixing structure for the leptons
 is naturally arisen in certain regions of the parameter space, moreover, the solar and
 atmospheric mass-squared differences for the neutrinos are also accommodated simultaneously.
 All the nontrivial results indicate clearly that the SUSY $SO(10)$ GUT model with the
 $\mathbf{120}$ Higgs couplings is worth investigating deeply for understanding the
 mystery of the fermion flavor.

\vspace{1cm}
 \noindent\textbf{Acknowledgments}

\vspace{0.3cm}
 One of the authors, W. M. Yang, would like to thank Hu Li for helpful discussions.
 This work is supported in part by National Natural Science Foundation of China.


\vspace{1cm}
\begin{thebibliography}{99}
\bibitem{1}
 For recent reviews, see H. Fritzsch and Z. Z. Xing, Prog. Part. Nucl. Phys. 45, 1 (2000);
 Z. Z. Xing, Int. J. Mod. Phys. A 19, 1 (2004);
 R. D. Mckeown and P. Vogel, Phys. Rept. 393, 315 (2004);
 A. Y. Smirnov, Int. J. Mod. Phys. A 19, 1180 (2004);
 R. N. Mohapatra, hep-ph/0402035;
 S. F. King and I. N. R. Peddie, hep-ph/0312235.
\bibitem{2}
 Y. Fukuda \emph{et al.} [Super-Kamiokande Collaboration], Phys. Rev. Lett. 81, 1562 (1998);
 Phys. Rev. Lett. 85, 3999 (2000);
 M. Apollonio \emph{et al.} [CHOOZ Collaboration], Phys. Lett. B 466, 415 (1999);
 Eur. Phys. J. C 27, 331 (2003);
 K. Eguchi \emph{et al.} [KamLAND Collaboration], Phys. Rev. Lett. 90, 021802 (2003);
 Q. R. Ahmad \emph{et al.} [SNO Collaboration], Phys. Rev. Lett. 89, 011301 (2002).
\bibitem{3}
 M. C. Gonzalez-Garcia and Y. Nir, Rev. Mod. Phys. 75, 345 (2003);
 P. C. de Holanda and A. Y. Smirnov, Astropart. Phys. 21, 287 (2004);
 Nucl. Phys. B680, 479 (2004);
 G. L. Fogli, E. Lisi, A. Marrone, D. Montanino, A. Palazzo, A. M. Rotunno,
 Phys. Rev. D 69, 017301 (2004);
 M. C. Gonzalez-Garcia and C. Pena-Garay, Phys. Rev. D 68, 093003 (2003).
\bibitem{4}
 G. Altarelli and F. Feruglio, hep-ph/0206077;
 S. F. King, Rept. Prog. Phys. 67, 107 (2004);
 S. Raby, Rept. Prog. Phys. 67, 755 (2004);
 J. C. Pati, Int. J. Mod. Phys. A 18, 4135 (2003);
 R. N. Mohapatra, \emph{Unification and Supersymmetry--The Frontiers of Quark-Lepton Physics},
 third edition, Springer-Verlag, New York (2003).
\bibitem{5}
 B. S. Babu and R. N. Mohapatra, Phys. Rev. Lett. 70, 2845 (1993);
 K. S. Babu and S. M. Barr, Phys. Rev. Lett. 75, 2088 (1995);
 C. H. Albright, Int. J. Mod. Phys. A 18, 3947 (2003);
 M. C. Chen and K. T. Mahanthappa, Int. J. Mod. Phys. A 18, 5819 (2003);
 M. Bando and M. Odara, Prog. Theor. Phys. 109, 995 (2003);
\bibitem{6}
 M. Gell-Mann, P. Ramond, R. Slansky, in \emph{Supergravity}, eds. P. van Niewenhuizen and
 D. Z. Freeman (North-Holland, Amsterdam, 1979);
 T. Yanagida, in \emph{Proc. of the Workshop on Unified Theory and Baryon Number in the Universe},
 eds. O. Sawada and A. Sugamoto, Tsukuba, Japan (1979);
 R. N. Mohapatra, G. Senjanovi\'{c}, Phys. Rev. Lett. 44, 912 (1980);
 R. N. Mohapatra and P. B. Pal, \emph{Massive Neutrinos in Physics and Astrophysics},
 third edition, World Scientific, Singapore (2004).
\bibitem{7a}
 G. G. Ross and L. Velasco-Sevilla, Nucl. Phys. B653, 3 (2003);
 S. F. King and G. G. Ross, Phys. Lett. B 574, 239 (2003);
 M. C. Chen and K. T. Mahanthappa, Phys. Rev. D 68, 017301 (2003);
 T. Asaka, Phys. Lett. B 562, 291 (2003);
 M. Bando and M. Obara, hep-ph/0307360.
\bibitem{7b}
 B. Bajc, A. Melfo, G. Senjanovic, F. Vissani, hep-ph/0402122;
 N. Oshimo, Phys. Rev. D 66: 095010 (2002).
\bibitem{8}
 D. G. Lee and R. N. Mohapatra, Phys. Rev. D 51, 1353 (1995);
 H. S. Goh, R. N. Mohapatra, S.-P. Ng, Phys. Lett. B 570, 215 (2003);
 Phys. Rev. D 68, 115008 (2003);
 T. Fukuyama and N. Okada, JHEP 0211, 011 (2002).
\bibitem{9}
 N. Oshimo, Nucl. Phys. B668, 258 (2003);
 C. S. Aulakh, B. Bajc, A. Melfo, G. Senjanovic, F. Vissani, Phys. Lett. B 588, 196 (2004);
 B. Bajc, hep-ph/0311214.
\bibitem{10}
 B. Bajc, G. senjanovi\'{c} and F. Vissani, Phys. Rev. Lett. 90, 051802 (2003);
 R. N. Mohapatra, hep-ph/0306016.
\bibitem{11}
 H. Georgi and C. Jarlskog, Phys. Lett. B 86, 297 (1979);
 H. Georgi and D. V. Nanopoulos, Nucl. Phys. B159, 16 (1979).
\bibitem{12}
 M. Kobayashi and T. Maskawa, Prog. Theor. Phys. 49, 652 (1973).
\bibitem{13}
 V. Barger, M. S. Berger and P. Ohmann, Phys. Rev. D 47, 1093 (1993);
 D. J. Castano, E. J. Piard and P. Ramond, Phys. Rev. D 49, 4882 (1994);
 K. S. Babu, C. N. Leung and J. Pantaleone, Phys. Lett. B 319, 191 (1993);
 S. Antusch, M. Drees, J. Kersten, M. Lindner and M. Ratz, Phys. Lett. B 525, 130 (2002).
\bibitem{14}
 B. M. Pontecorvo, Sov. Phys. JETP 6, 429 (1958);
 Z. Maki, M. Nakagawa and S. Sakata, Prog. Theor. Phys. 28, 870 (1962).
\bibitem{15}
 K. Hagiware \emph{et al.} [Particle Data Group], Phys. Rev. D 66, 010001 (2002).
\end{thebibliography}
\end{document}